\documentclass[%
 reprint,
 amsmath,amssymb,
 aps,
]{revtex4-2}

\usepackage{graphicx}
\usepackage{dcolumn}
\usepackage{bm}
\usepackage{subcaption}

\begin{document}

\preprint{APS/123-QED}

\title{An extension of Lambertson method to horizontally asymmetric Beam Position Monitors}

\author{T. Luo}
\email{tluo@lbl.gov}
\author{C. Sun}
\author{J. De Chant}
\author{A. Zabala}
\affiliation{Lawrence Berkeley National Laboratory, Berkeley 94720, California}

\date{\today}

\begin{abstract}
Lambertson method is a classical approach that can indirectly measure the electrical center offset in the beam position monitor (BPM) due to imperfections in the pickup buttons. It applies to BPMs that are symmetric in both the horizontal and vertical directions. In this paper, we present our extension of this method to BPMs that are asymmetric in the horizontal direction, and we apply it to such BPMs in ALS Upgrade project at Lawrence Berkeley National Laboratory.  

\end{abstract}

\maketitle

\section{Introduction}

Lambertson method~\cite{lambertson} is a classical approach for indirectly measuring electrical center offsets in a beam position monitor (BPM) arising from the imperfect BPM pickups. It is applicable to BPMs with both horizontal and vertical geometrical symmetry.    

However, in modern synchrotron light sources such as ALS Upagrade (ALS-U)~\cite{Steier:2019sbc}, due to the limited beamline space, some BPMs have to share the real estate with other components such as antechambers. Thus the horizontal symmetry is broken due to an opening on one side of the BPM chamber. Therefore, an extension of the Lambertson method is required to apply it to horizontally asymmetric BPMs.  

\section{Review of the Lambertson method}

For an ideal BPM, when the beam passes through the electric center, the signal induced on the four pickups are equal, and the BPM position readings are zero. In practice, the pickup imperfections, such as insertion depth, button capacitance, etc.) are unavoidable and the gain of each pickup is not exactly equal. Even if the beam is located at the electric center, the uneven signals from the four pickups can produce an artificial non-zero position reading. Calibrating out these artificial offsets is critical for obtaining an accurate absolute beam position, which is particularly important for the initial closed-orbit commissioning.

Before installed on the beamline, the electrical center offsets of the BPM blocks can be measured directly by the wire methods. In this method, a metallic wire passes through the BPM center, carrying a TEM wave to mimic the beam EM field below the cutoff frequency of the first TE/TM mode. The beam induced signals can be measured at each pickup from which one can calculate the offsets. This method requires a high accuracy of the wire positioning, as well as a good impedance matching through the path of the TEM wave.

\begin{figure}[htbp]
\includegraphics[width=0.8\linewidth]{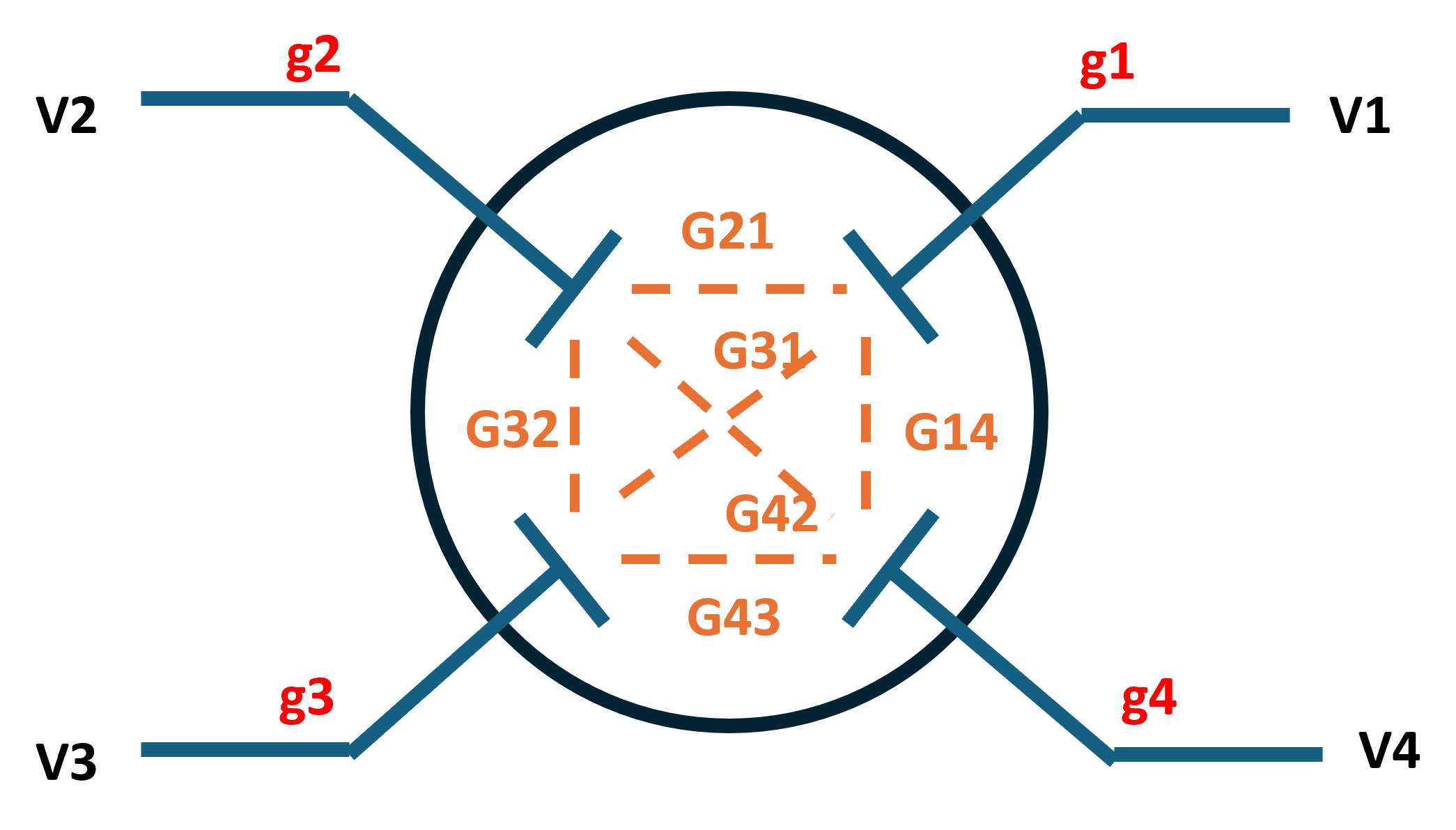}
\caption{\label{fig:sym-bpm-illustration} A symmetric BPM with coupling between each pickup $G_{ij}$ and the gain at each pickup $g_k$.}
\end{figure}

By contrast, Lambertson method is an indirect way to measure the offsets without extra setups. It can be carried out both before or after the BPM block is installed on the beamline. We used this method to measure the electric center offsets for the ALS-U Accumulate Ring BPMs~\cite{luo:ipac2024-wepg61}. In this method, one measures the transmission coefficient between each pickup, with the coefficient from pickup $n$ to pickup $m$ denoted as $S_{mn}$. Assuming the coupling between pickup $i$ and $j$ in the vacuum chamber is $G_{ij}$, and the signal gain of each pickup $k$ is $g_k$, as illustrated in Fig.~\ref{fig:sym-bpm-illustration}, we have:
\begin{eqnarray*}
    S_{21}=g_1\cdot G_{21}\cdot g_2 \\
    S_{23}=g_3 \cdot G_{23} \cdot g_2 \\
    S_{24}=g_4 \cdot G_{24} \cdot g_2 \\
    S_{31}=g_1 \cdot G_{31} \cdot g_3 \\
    S_{34}=g_4 \cdot G_{34} \cdot g_3 \\
    S_{41}=g_1 \cdot G_{41} \cdot g_4
\end{eqnarray*}

For a BPM with both horizontal and vertical symmetry, the couplings between the pickups satisfy:
\begin{eqnarray*}
    G_{12}=G_{12}=G_{34}=G_{43} \\
    G_{23}=G_{32}=G_{14}=G_{41}\\
    G_{13}=G_{31}=G_{24}=G_{42}
\end{eqnarray*}

With these relations, the pick up gains can be written as:
\begin{eqnarray*}
    g_1^2=\frac{S_{21}S_{14}}{S_{42}}K_A=\frac{S_{12}S_{31}}{S_{32}}K_B=\frac{S_{41}S_{13}}{S_{43}}K_C \\
    g_2^2=\frac{S_{21}S_{32}}{S_{31}}K_A=\frac{S_{12}S_{42}}{S_{14}}K_B=\frac{S_{32}S_{42}}{S_{43}}K_C \\
    g_3^2=\frac{S_{32}S_{43}}{S_{42}}K_A=\frac{S_{43}S_{31}}{S_{14}}K_B=\frac{S_{32}S_{13}}{S_{21}}K_C \\
    g_4^2=\frac{S_{14}S_{43}}{S_{13}}K_A=\frac{S_{43}S_{24}}{S_{32}}K_B=\frac{S_{14}S_{42}}{S_{21}}K_C 
\end{eqnarray*}
, where:
\begin{eqnarray*}
    K_A=\frac{G_{13}}{G_{12}G_{23}} \\
    K_B=\frac{G_{23}}{G_{12}G_{13}} \\
    K_C=\frac{G_{12}}{G_{23}G_{13}} 
\end{eqnarray*}

When a beam passes through the electric center, the signal it induces on each pickup is proportional to $g_k$. For an ideal BPM, all $g_k$ should be equal. The electrical and mechanical imperfection of the pickups, such as the button insertion depth, the button capacitance, etc., could cause a variation in $g_k$ and thus cause an artificial center offset. The horizontal and vertical offset due to $g_k$ mismatch can be calculated as:
\begin{eqnarray*}
    dX=\frac{1}{S_x}\frac{g_1+g_4-g_2-g_3}{g_1+g_2+g_3+g_4} \\
    dY=\frac{1}{S_y}\frac{g_1+g_2-g_3-g_4}{g_1+g_2+g_3+g_4}
\end{eqnarray*}
, where $S_{x}$ and $S_y$ are the horizontal and vertical BPM sensitivity respectively. The sensitivity represents the linear correlation between the beam position offset and the normalized differential BPM signal. 

Without knowing $K_{A/B/C}$, we can still calculate $dX$ or $dY$ as long as we use the terms with the same $K$ factor for calculating $g_i$. This is the core feature of the Lambertson method.

If the pickups are adjustable, the measured $g_i$ value can also guide how to adjust them to reduce the offsets. For example, if one $g_i$ is significantly smaller than other three, this pickup can be inserted further to increase its $g_i$ to balance with other three.  

It should be emphasized that one essential assumption of the Lambertson method is that the BPM is geometrically symmetric in both horizontal and vertical direction.

\section{Horizontally asymmetric BPMs in modern particle accelerators}

Ideally, the BPM chamber should be geometrically symmetric in both horizontal and vertical directions. Common shapes of BPM chamber include circular, elliptical, rectangular, diamond-like, etc. 

\begin{figure}[htbp]
\includegraphics[width=0.8\linewidth]{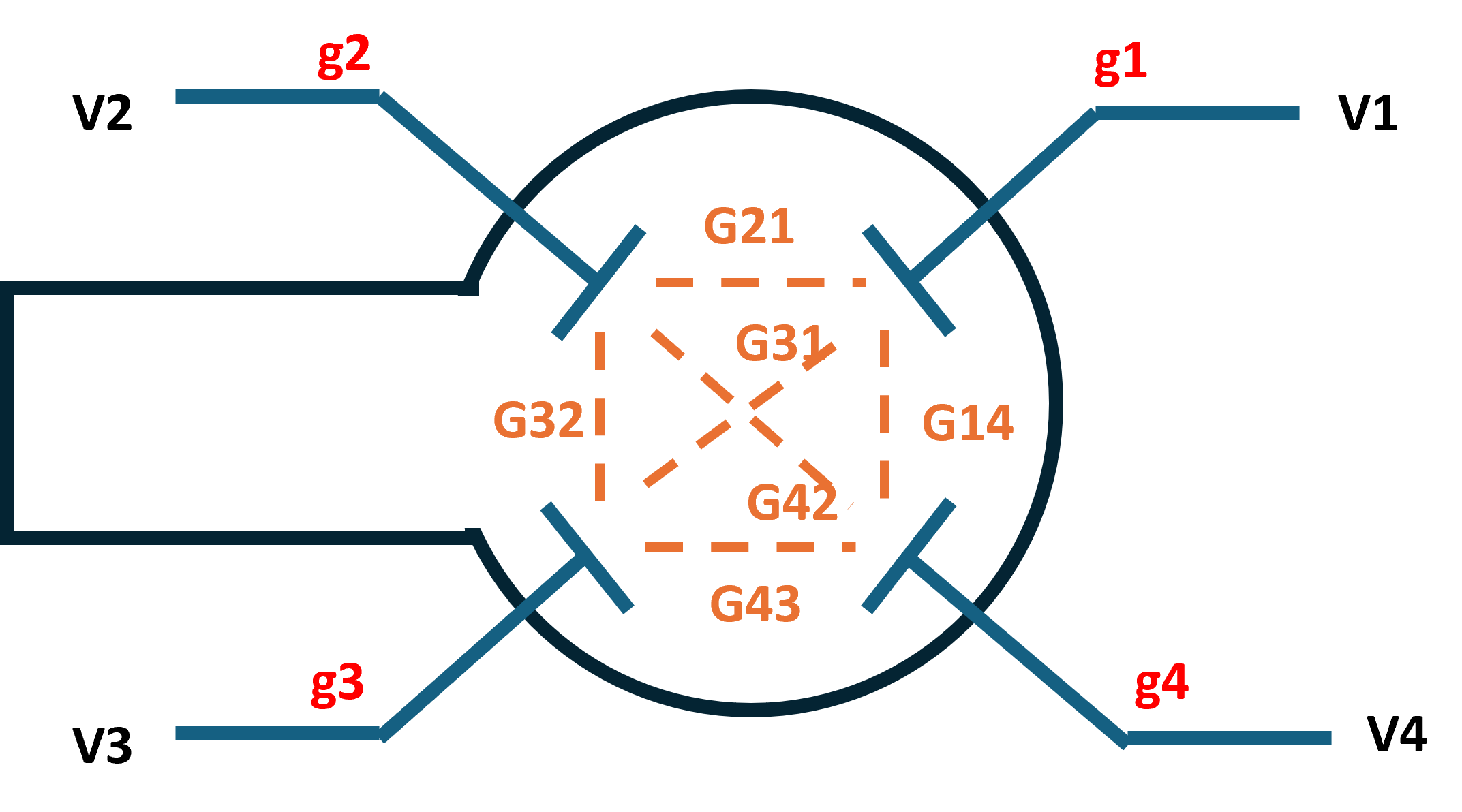}
\caption{\label{fig:asym-bpm-illustration} An asymmetric BPM with coupling between each pickup $G_{ij}$ and the gain at each pickup $g_k$.}
\end{figure}

In modern accelerators, due to the limited beamline space, the BPM may have to share the intallation space with other components, such as the synchrotron light ante-chamber. Such structures are usually only on one side of the BPM and therefore break the horizontal symmetry, while leaving only the vertical symmetry intact, as shown in Fig~\ref{fig:asym-bpm-illustration}.  

\begin{figure}[htbp]
\includegraphics[width=0.8\linewidth]{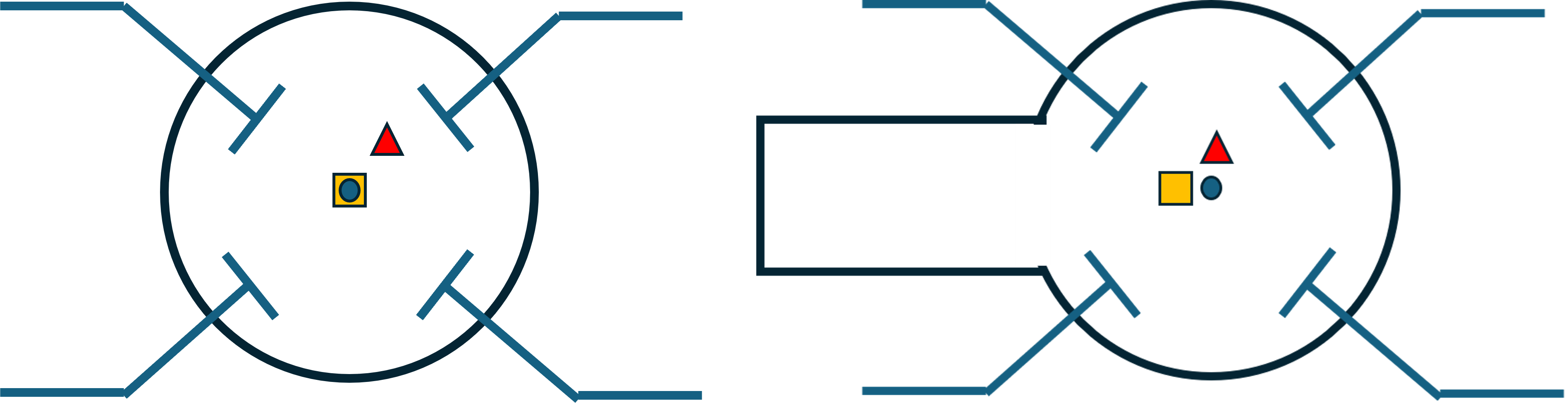}
\caption{\label{fig:bpm-center-illustration} In a symmetric BPM, the chamber geometric center (blue dot) overlaps with the electric center (yellow square). In an asymmetric BPM, the chamber geometric center (ignore the antechamber) is no longer overlap with the electric center. The extended Lambertson method presented in this paper calculates the offsets relative to the electric center for the asymmetric BPM shown on the right.}
\end{figure}

Like the symmetry BPM, we need to characterize the sensitivity $S_x$ and $S_y$ of the asymmetric BPM. In additional, its electric center will have a horizontal offset $o_x$ from the chamber geometric center, which is often designed to coincide, or be very close, to the closed orbit. This offset needed to be characterized as well. A comparison of the geometric center, electric center and the electric center offset are illustrated in Fig~\ref{fig:bpm-center-illustration} for both the symmetric an asymmetric BPM. 

Here we present how we calculate the sensitivity $S_x$, $S_y$ and the offset $o_x$ for the horizontally asymmetric BPMs in ALS-U Storage Ring. In CST~\cite{CST} particle solver, a beam passed through a set of grid point $(x,y)$ around the geometric center, which is the origin of the coordinate system. The grid points lie within a radius of 2\,mm from the geometric center. At each location, the signal in time domain from each pickup $i$ is recorded as $P_i(x,y)$. Then we can calculate its frequency spectrum by Fourier transform, and get the signal strength $F_i(x,y)$ at the chosen frequency in electronics, which is the RF frequency of $f_0=500$\,MHz for ALS-U. For each beam passing through one grid, we compute:

\begin{eqnarray*}
    X_{cal} &=&\frac{1}{S_x}\frac{F_1+F_4-F_2-F_3}{F_1+F_2+F_3+F_4} + o_x \\
    Y_{cal} &=& \frac{1}{S_y}\frac{F_1+F_2-F_3-F_4}{F_1+F_2+F_3+F_4}
\end{eqnarray*}

 For $n$ grid points, we obtain $n$ set of such equations. $S_x$, $S_y$ and $o_x$ can be determined by the linear regression, following the rule that we want to maximize the number of grid points whose non-linear position errors are within the acceptable level, which is 0.1\,mm for ALS-U:
 \begin{eqnarray*}
     |X_{cal}-x|\le0.1\text{\, mm} \\
     |Y_{cal}-y|\le0.1\text{\, mm}
 \end{eqnarray*}

For some BPMs, there are multiple values of $S_x$, $S_y$ and $o_x$ that can have all the grid points satisfying the position error requirements above. For such cases, we choose the median value as the best fit.

\section{Extension of Lambertson method for BPMs without horizontal symmetry}
When only vertical symmetry is present, we still have 
\begin{eqnarray*}
    G_{12}=G_{21}=G_{34}=G_{43} \\
    G_{13}=G_{31}=G_{24}=G_{42}
\end{eqnarray*}
However, the lack of horizontal symmetry implies:
\begin{eqnarray*}
    G_{23}=G_{32}\neq G_{14}=G_{41} 
\end{eqnarray*}
Therefore, the classical Lambertson method is no longer valid. We follow the same paradigm of deriving Lambertson method, but introduce a new factor $\eta$ to represent the horizontal asymmetry:

\begin{eqnarray*}
     G_{41}=\eta \cdot G_{23}
\end{eqnarray*}

With $\eta$, the pickup gains can be rewritten in an analogous form:

\begin{eqnarray*}
    g_1^2 &=&\frac{1}{\eta}\frac{S_{21}S_{14}}{S_{42}}K_A=\frac{S_{12}S_{31}}{S_{32}}K_B=\frac{1}{\eta}\frac{S_{41}S_{13}}{S_{43}}K_C \\
    g_2^2  &=&\frac{S_{21}S_{32}}{S_{31}}K_A=\eta\frac{S_{12}S_{42}}{S_{14}}K_B=\frac{S_{32}S_{42}}{S_{43}}K_C \\
    g_3^2 &=&\frac{S_{32}S_{43}}{S_{42}}K_A=\eta \frac{S_{43}S_{31}}{S_{14}}K_B=\frac{S_{32}S_{13}}{S_{21}}K_C \\
    g_4^2 &=&\frac{1}{\eta}\frac{S_{14}S_{43}}{S_{13}}K_A=\frac{S_{43}S_{24}}{S_{32}}K_B=\frac{1}{\eta}\frac{S_{14}S_{42}}{S_{21}}K_C 
\end{eqnarray*}

Now in order to solve $g_i$ without knowing $K_j$, we need to know this extra factor $\eta$. For a known BPM geometry, $\eta$ can be solved by numerically simulating the transmission coefficients $S_{41}$ and $S_{23}$ for a BPM with four perfect pickups. In such a BPM, the gain of each pickup $g_i$ are all equal. Thus:

\begin{equation*}
    \eta=\frac{G_{41}}{G_{23}}=\frac{S_{41}}{S_{23}}
\end{equation*}

Here the underlying assumption is that the imperfect pickups can cause the electric center offsets but has negligible effect on $\eta$, thus $eta$ can be treated as a constant.  

\section{Validating with CST simulation}

Numerical simulation is carried out to validate this method. The validating procedure is:
\begin{enumerate}
    \item {Determine the sensitivity $S_x$ and $S_y$, and the electric center relative to the geometric center $o_x$ for a BPM with perfect pickups, as described in Sec.III.}
    \item {Simulate the transmission coefficients for this BPM and calculate the $\eta$}
    \item {Introduce the pickup gain errors by inserting or recessing the pickups.}
    \item {Simulate a beam passing through the geometric center of this imperfect BPM in CST Particle Studio. From the beam-induced signal on each pickup $P_i$, one can calculate its frequency spectrum through Fourier transform. In this spectrum, the signal strength at 500 MHz is $F_i$. The electric center offsets caused by pickup errors are calculated as:
    \begin{eqnarray*}
        dX_{PT}&=&\frac{1}{S_x}\frac{F_1+F_3-F_2-F_4}{F_1+F_2+F_3+F_4} +o_x\\
        dY_{PT}&=&\frac{1}{S_y}\frac{F_1+F_2-F_3-F_4}{F_1+F_2+F_3+F_4}
    \end{eqnarray*}
    }
    \item{For the same imperfect BPM, simulate the transmission coefficient $S_{ij}$ through each pickup pairs. From $S_{ij}$ and $\eta$, derive $g_i$ and calculate the center offsets $dX_{EL}$ and $dY_{EL}$ by the extended Lambertson (EL) method. Agreement between $dX_{EL}$/$dY_{EL}$ and $dX_{PT}$/$dY_{PT}$ validates the extended Lambertson method.}
\end{enumerate}

\begin{figure}[htbp]
\includegraphics[width=0.8\linewidth]{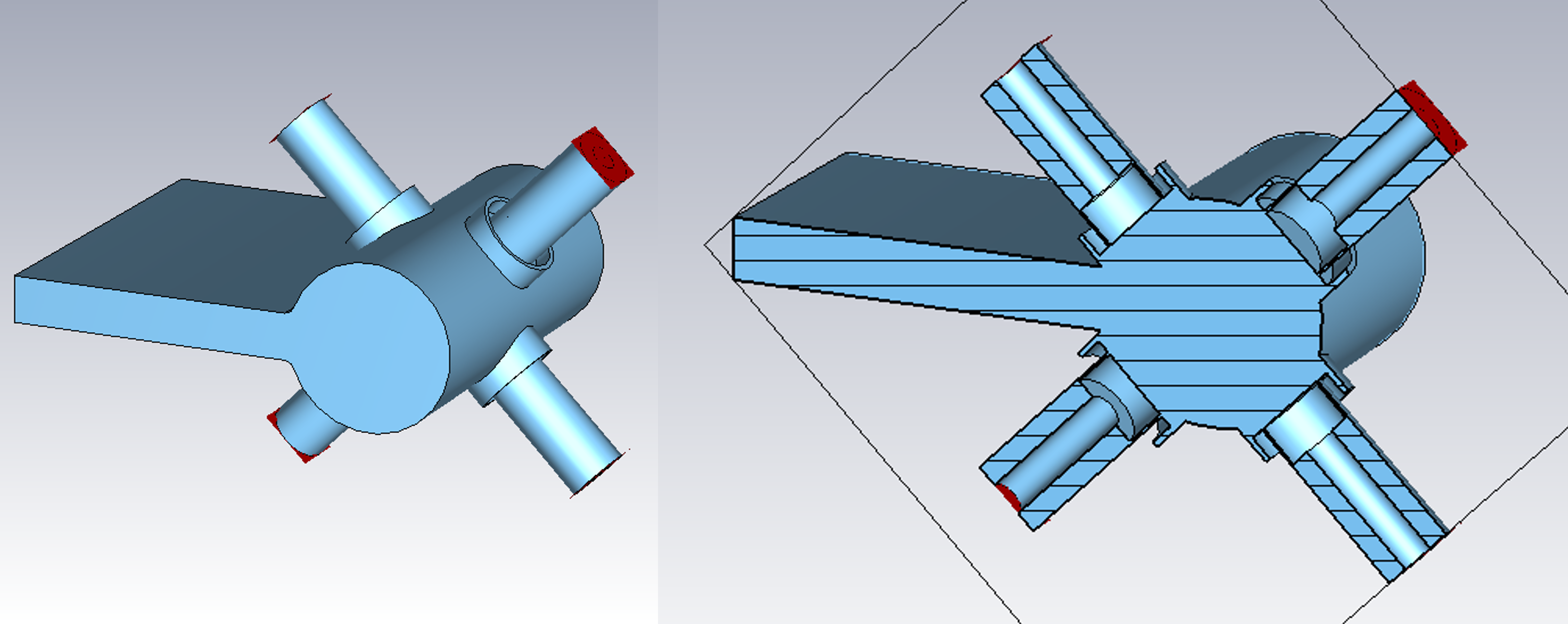}
\caption{\label{fig:bpm_f2} The asymmetric BPM F2 in ALS-U.}
\end{figure}

\begin{figure}[htbp]
\includegraphics[width=0.8\linewidth]{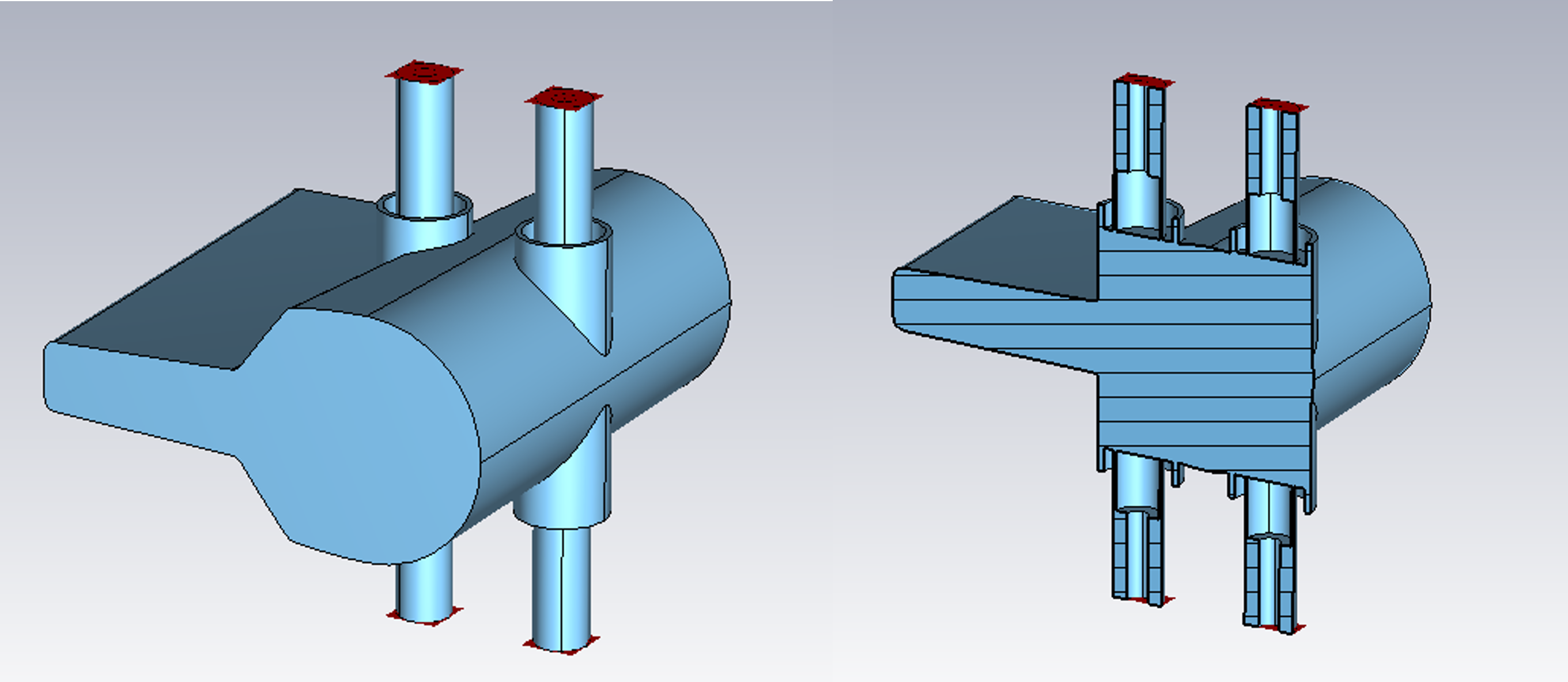}
\caption{\label{fig:bpm_j1} The asymmetric BPM J1 in ALS-U.}
\end{figure}

Two ALS-U Storage Ring BPMs, labeled as type F2 and J1, are used for validation. F2 is a round BPM with a 18\,mm chamber diameter, a 6\,mm button diameter, and a 5\,mm keyhole width, as shown in Fig.~\ref{fig:bpm_f2}. J1 is more asymmetric and irregular than F2, with a tapered chamber diameter from 18 to 20\,mm, a 6\,mm button diameter, and a tapered keyhole shape, as shown in Fig.~\ref{fig:bpm_j1}.

Following the procedure described in Sec.III, we calculated the $S_x$, $S_y$ and $o_x$ for both BPMs. We also simulated the asymmetric factor $\eta$ from the s parameters as desribed in Sec.VI. The results are shown in Table~\ref{tab:bpm_parameters}

\begin{table}[htbp]
\caption{\label{tab:bpm_parameters}%
The sensitivities and horizontal offset for ALS-U BPM F2 and J1 
}
\begin{ruledtabular}
\begin{tabular}{lllll}
\textrm{BPM type}&
\textrm{$S_x$ (\%/mm)}&
\textrm{$S_y$ (\%/mm)}&
\textrm{$o_x$ (um)}&
\textrm{$\eta$}\\
\colrule
F2 & 15.3 & 15.0 & 149 & 0.83\\
J1 & 12.1 & 15.8 & -379 & 0.89\\
\end{tabular}
\end{ruledtabular}
\end{table}

\begin{figure}[htbp!]
    \centering
        \centering
        \includegraphics[width=0.45\textwidth]{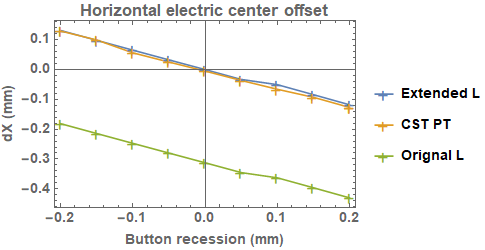}   
    \vspace{1em} 
        \centering
        \includegraphics[width=0.45\textwidth]{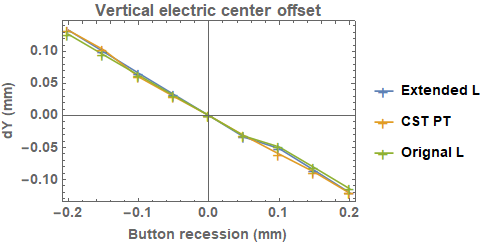}       
    \caption{Center offsets comparison for BPM F2}
    \label{fig:f2_comparison}
\end{figure}

\begin{figure}[htbp!]
    \centering
        \centering
        \includegraphics[width=0.45\textwidth]{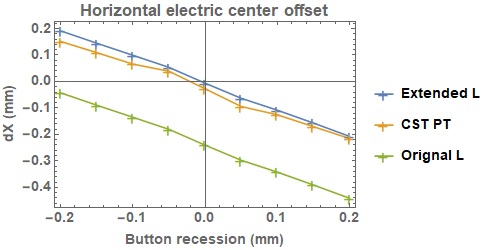}   
    \vspace{1em} 
        \centering
        \includegraphics[width=0.45\textwidth]{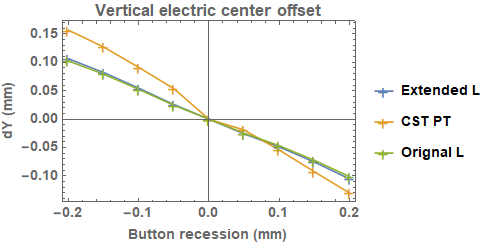} 
    \caption{Center offsets comparison for BPM J1}
    \label{fig:j1_comparison}
\end{figure}

To implement an error on the BPM pickups, we shift the position of the upper left button from +200\,um insertion to -200\,um recession with a 50\,um step size. $dX_{EL}$/$dY_{EL}$ and $dX_{PT}$/$dY_{PT}$ are calculated for each button position. The results are shown in Figure~\ref{fig:f2_comparison} for BPM F2 and Figure~\ref{fig:j1_comparison} for BPM J1. For comparison, the offsets calculated using the original (classical) Lambertson method are also included. 

These results show that:
\begin{enumerate}
    \item {Overall, the extended Lambertson methods agrees well with the particle tracking simulations.}
    \item {For the vertical offsets of BPM J1, there is a larger discrepancy than F2. But the differences remain within 50\,um which is acceptable. This may be due to that compared with BPM F2, BPM J1 has a larger offset between the electric center and the geometric center, as well as a stronger non-linear relation between the beam position and the BPM signals around the center.}
    \item{If the classical Labmertson method is applied directly without the factor $\eta$, there will be a large horizontal offset error. It still works for the vertical offset because vertical symmetry is preserved.}
\end{enumerate}

\section{Application to ALS-U asymmetric BPMs} 

We are applying this extended Lambertson method to measure the electric center offsets for ALS-U asymmetric BPM productions. Here we showed the measurement of two Storage Ring BPMs: the first assembly of F2 block and the first assembly J1 block. 

\begin{figure}[htbp]
\includegraphics[width=0.95\linewidth]{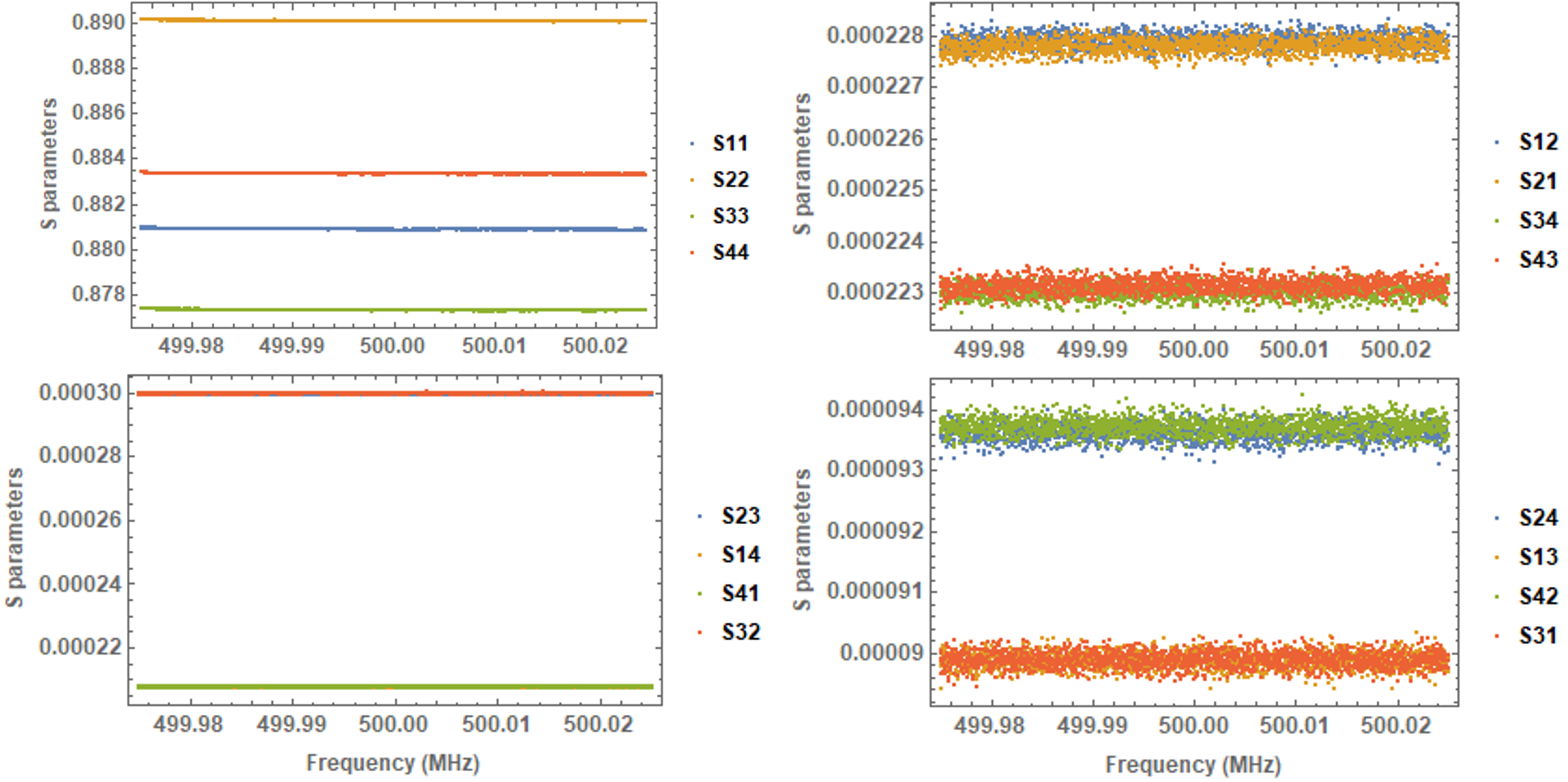}
\caption{\label{fig:s-f2} The s parameters measurement of BPM F2.}
\end{figure}

\begin{figure}[htbp]
\includegraphics[width=0.95\linewidth]{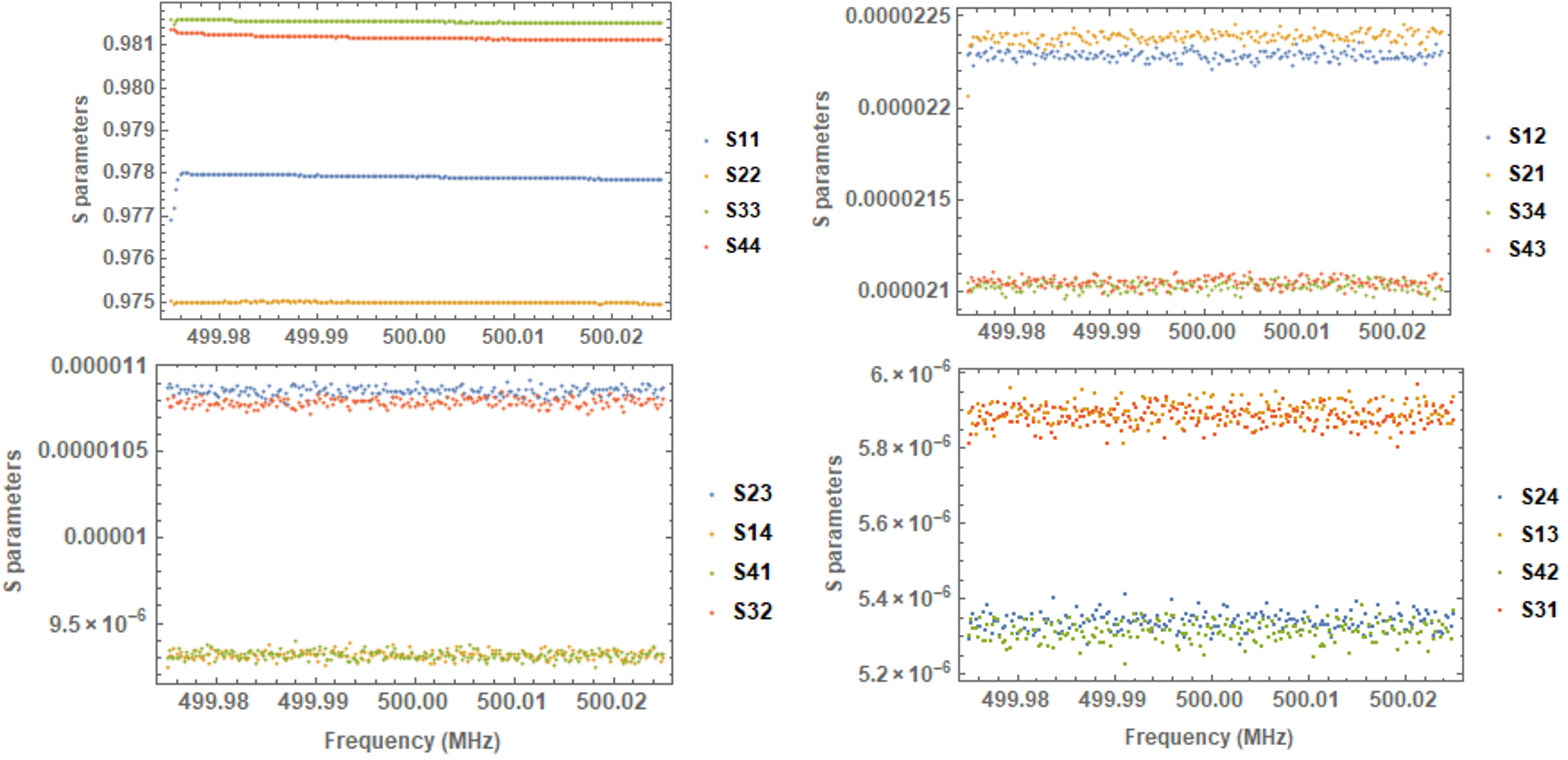}
\caption{\label{fig:s-j1} The s parameters measurement of BPM J1.}
\end{figure}

The s parameters are measured using a 4-port Network Analyzer (NWA) as shown in Fig.~\ref{fig:s-f2} and Fig.~\ref{fig:s-j1}. Note that couplings between the pickups can be weak, resulting in small values of s parameters. To achieve sufficient accuracy, the NWA output power is set at the maximum level and data is averaged over multiple data-takings. For an ideal BPM with horizontal asymmetry and vertical symmetry, we expect:

\begin{eqnarray*}
 \frac{S_{12}}{S_{21}}=\frac{S_{34}}{S_{43}},\quad \frac{S_{24}}{S_{42}}=\frac{S_{13}}{S_{31}}   
\end{eqnarray*}

 For a real BPM with pickup errors, these ratios are not equal, as shown in the right two plots in Fig.~\ref{fig:s-f2} and Fig.~\ref{fig:s-j1}. Their differences provides the exact information to deduce the gain differences at each pickup in the extended Lambertson method.

Using the measured s parameters around 500 MHz, the center offsets calculated by the extended Lambertson method are listed in Table.~\ref{tab:measured_offsets}. The results show that this F2 block has a fairly large horizontal offset of 288\,um. This large offset is consistent with the mechanical measurement results from the chamber vendor, and will be corrected by adding shims beneath the button flanges to adjust the button insertion depths.   

\begin{table}[htbp]
\caption{\label{tab:measured_offsets}%
The center offsets measurement of the first assembled F2 and J1 
}
\begin{ruledtabular}
\begin{tabular}{lll}
\textrm{BPM type}&
\textrm{$d_x$ (um)}&
\textrm{$d_y$ (um)}\\
\colrule
F2 & 288 & -25\\
J1 & 69 & -93\\
\end{tabular}
\end{ruledtabular}
\end{table}

\section{Summary} The classical Lambertson method can indirectly calculate the BPM electric centr offsets caused by imperfect pickups. It is only valid when the BPM is geometrically symmetric in both horizontal and vertical direction. In this paper, we presented an extension to the classical Lambertson method to calculate the electric center offsets for the horizontal asymmetric BPMs. The key is to introduce a new factor $\eta$ to account for the asymmetric transmissions between the pickup pair on the left and the right side of BPM. We validate this new approach using numerical simulations of two asymmetric ALS-U BPMs. Currently we are applying this method to the asymmetric ALS-U Storage Ring BPMs under production.

\begin{acknowledgments}
We thank Mouda Tang, Sol Omolayo, Matthew McHenry and Raul Mascote on the BPM bench test, as well as Mike Chin, Christoph Steier and Marco Verturini on the BPM electronics and beam physics discussion. This work is supported by Director of Science of the U.S. Department of Energy under Contract No. DE-AC02-05CH11231. 
\end{acknowledgments}

\bibliography{reference}

\end{document}